\documentclass[useAMS,usenatbib]{mn2e}
\usepackage{graphicx}
\usepackage{subfigure}
\def\bmu{\mbox{\boldmath $\mu$}}
\title[Joint Bayesian separation and restoration]{Joint Bayesian separation and restoration of CMB from convolutional mixtures}
\author[Kayabol et al.]{K. Kayabol$^{1,3}$\thanks{E-mail:
koray.kayabol@inria.fr (KK); sanz@ifca.unican.es (JLS);
herranz@ifca.unican.es (DH); ercan.kuruoglu@isti.cnr.it (EEK);
emanuele.salerno@isti.cnr.it (ES)}, J. L.
˜Sanz$^{2}$\footnotemark[1], D. ˜Herranz$^{2}$\footnotemark[1], E.
E. ˜Kuruoglu$^{1}$\footnotemark[1] and E.
˜Salerno$^{1}$\footnotemark[1]
\\
$^{1}$ISTI, CNR, via G. Moruzzi 1, 56124, Pisa, Italy\\
$^{2}$IFCA, University of Cantabria, Avda. Los Castros s/n 39005,
Santander, Spain\\
$^{3}$Project-Team Ariana, INRIA, 2004 route des Lucioles, 06902
Sophia Antipolis, France}
\begin{document}

\date{Accepted 1988 December 15. Received 1988 December 14; in original form 2010 July 16}

\pagerange{\pageref{firstpage}--\pageref{lastpage}} \pubyear{2010}

\maketitle

\label{firstpage}

\begin{abstract}
We propose a Bayesian approach to joint source separation and
restoration for astrophysical diffuse sources. We constitute a
prior statistical model for the source images by using their
gradient maps. We assume a $t$-distribution for the gradient maps
in different directions, because it is able to fit both smooth and
sparse data. A Monte Carlo technique, called Langevin sampler, is
used to estimate the source images and all the model parameters
are estimated by using deterministic techniques.
\end{abstract}


\section{Introduction}

Inferring the CMB radiation map is an important task to estimate
the cosmological parameters. The foreground radiation
contamination at related observation frequencies, the noise
degradation of the instruments and the blur caused by the antenna
apertures make this task very difficult. Under Independent
Component Analysis (ICA) framework, separation of the CMB
radiation among the others has been done by \citet{Maino02}. In
\citep{Cardoso02,Bedini05,Bonaldi07}, the noise has been taken
into consideration to find the separation matrix and the maps are
obtained by using generalized Least Square (LS) solution.
\citet{Wilson08}, \citet{Eriksen08} and \citet{Kayabol09ip} have
used Bayesian approach for separation and noise removal of the
maps. The point spread functions of the antennas are included in
\citet{Bedini07} and \citet{Ricciardi10} to estimate a parametric
mixing matrix, but they are not considered in the map
reconstruction process.

In this study, we focus on the problem of multi-channel source
separation and restoration from multi-channel blurred and noisy
observations with channel-variant point spread functions (psf).
The resolutions of the observed channel maps are generally
different, since the aperture of the telescope beam depends on
frequency. We perform the source separation, the de-noising and
the de-blurring processes together. By considering the previous
studies \citep{Bonaldi07, Ricciardi10}, we assume that the
non-linear parameters of the mixing matrix are known with an
error. Under this assumption, we reconstruct the source maps in
the pixel domain by using a Monte Carlo technique that has been
recently developed and tested on the astrophysical source
separation problem \citep{Kayabol10}. Our method is an extended
version of the method in \citep{Kayabol10} to convolutional
mixture problem and has also the ability to estimate the mixing
matrix.

Studies on separation of convolutional or blurred image mixtures
can be found in the image processing literature.
\citet{Castella04} extended the contrast function based ICA
technique in the case of blurring. \citet{Anthoine05} proposed to
solve the same problem by adapting the existing variational and
statistical methods and modeling the components in wavelet domain.
\citet{Tonazzini05} use the Markov Random Field (MRF) based image
prior in the Bayesian framework. \citet{Shwartz08} address a
solution to separation of defocus blurred reflections in the
natural scenes by using the sparsity of the Short Time Fourier
Transform (STFT) coefficients as priors. In a recent study
\citep{Tonazzini10}, multi-channel separation and deconvolution is
proposed for document images. We use a Bayesian formulation to
include the effects of the antenna apertures and solve the
deblurring and map reconstruction problem jointly. Since the psf's
of the antennas are known, we easily define our likelihood
function by resorting to them.

In a Bayesian framework, we define prior densities for the source
maps. Because of the blur and the noise, the reconstruction
problem is very badly conditioned. It means that we have already
lost some detail information on the observed image. The lost
information in such a case is found in the high frequency contents
of the images. While choosing our image prior, we consider this
situation and define a prior that models the distribution of the
high frequency components of the image. We use the most basic high
frequency components of the image, namely image differentials. We
obtain the image differentials by applying a simple horizontal and
vertical gradient operator. The intensities of the image
differentials are very sparse and have a heavy-tailed
distribution. We exploit the $t$-distribution as a statistical
model for the image differentials. The first examples of use of
the $t$-distribution in inverse imaging problems can be found in
\citep{Higdon94,Prudyus01}. In \citep{Prudyus01}, it is reported
that the $t$-distribution approximates accurately the wavelet
coefficients of an image. In recent papers, it has been used for
image restoration \citep{Chantas08} and deconvolution
\citep{Tzikas09}.

In \citep{Kayabol10}, it is empirically shown that the image
differentials of the CMB, synchrotron and dust maps can be modeled
by $t$-distribution, which can be then used in Bayesian source
separation. Since the CMB is assumed to be a Gaussian random
field, the image differentials of CMB is more smooth than the
other components. Its differential might be modelled as a
Gaussian, but in this study we model it as a $t$-distribution by
using the fact that the $t$-distribution approaches to a Gaussian,
if its degree of freedom (dof) parameter goes to infinity. In
computer experiments, we can deal with infinity by replacing it by
large numbers.

Using the statistics of the image differentials as a prior for
separation and reconstruction does not introduce new information
into the data, but emphasizes some part of the data to help the
solution of the problem. The important part of the Bayesian image
reconstruction problem is determining the contribution of the
prior to the solution. If it is defined by the user, the
expectations of the user might be introduced into the solution. It
can be useful for natural, photographic and medical images that
are enhanced by the user, but in the case of astrophysical images,
since some of the physical parameters will be estimated after
reconstruction, the contribution of the prior must be controlled
automatically. In this study, the dof parameter of the
$t$-distribution controls the contribution of the prior to the
solution, and we estimate this parameter from data along the
iterations. The dispersion (scale) parameter of the
$t$-distribution is also estimated in the algorithm.

The organization of the paper is as follows. We introduce the
astrophysical component separation problem in the case of
convolutional mixtures in Section \ref{astrocomp}. In Section
\ref{probdef}, we define formally the source separation problem in
the Bayesian context, and outline the source model, the likelihood
and the posteriors. The details of the source maps and parameters
estimations are given in Section \ref{algo}. A number of
simulation cases including for five different sky patches are
given in Section \ref{simres}, and finally conclusions are drawn
in Section \ref{conc}.

\section{Component Separation Problem: Convolutional Mixtures Case} \label{astrocomp}

Let the $k$th observed pixel be denoted by $y_{k,i}$, where $i \in
\{ 1,2,\ldots,N \}$  represents the lexicographically ordered
pixel index. We assume that the observed images, $\mathbf{y}_{k}$,
$k \in \{1,2,\ldots,K\}$, are some linear combinations of source
images, $\mathbf{s}_{l}$, $l \in \{1,2,\ldots,L\}$. Taking into
account the effect of the telescope, the observation model can be
written as

\begin{equation}\label{mix2}
    \mathbf{y}_{k} = \mathbf{h}_{k}\ast\sum_{l=1}^{L}a_{k,l}\mathbf{s}_{l} + \mathbf{n}_{k}
\end{equation}
where the asterisk means convolution, and $\mathbf{h}_{k}$ is the
channel-variant telescope point spread function (psf) in the
$k$'th observation channel here assumed as Gaussian and circularly
symmetric. The observation model is not an instantaneous linear
mixing, since $\mathbf{h}_{k}$ changes for each channel. The
vector $\mathbf{n}_{k}$ represents an iid zero-mean Gaussian noise
with $\Sigma = \sigma_{k}^{2} \mathbf{I}_{N}$ covariance matrix
where $\mathbf{I}_{N}$ is an identity matrix. Although the noise
is not homogeneous in the astrophysical maps, we assume that the
noise variance is homogeneous within each sky patch and is also
known.

\section{Bayesian Formulation of Astrophysical Component Separation} \label{probdef}

\subsection{Source Model}

We used the self similarity based image model previously proposed
in \citep{Kayabol10}. In this model, we assume that the
intensities of the neighboring pixels are closed each other. To
express a pixel by using its neighbors, we write an
auto-regressive source model using the first order neighbors of
the pixel in the direction $d$:

\begin{equation}\label{AR}
    \mathbf{s}_{l} = \alpha_{l,d}\mathbf{G}_{d}\mathbf{s}_{l} + \mathbf{t}_{l,d}
\end{equation}
where the maximum number of first order neighbors is 8 but we use
only 4 neighbors, $d \in \{1,\ldots, 4\}$, in the main vertical
and horizontal directions. The matrix $\mathbf{G}_{d}$ is a linear
one-pixel shift operator, $\alpha_{l,d}$ is the regression
coefficient and the regression error $\mathbf{t}_{l,d}$ is an iid
$t$-distributed zero-mean vector with dof parameter $\beta_{l,d}$
and scale parameters $\delta_{l,d}$. To penalize the large
regression error occurred in the sharp edge regions of the image,
we use the $t$-distribution. Generally in real images, except the
Gaussian distributed ones, the regression error is better modelled
by some heavy-tailed distribution. The $t$-distribution can also
model the Gaussian distributed data. Therefore it is a convenient
model for data whose distribution ranges from Cauchy to Gaussian.
In \citep{Kayabol10}, $t$-distribution has been fitted to
simulated CMB, synchrotron and dust maps and gives better results
in the sense of mean square error when compared to Gaussian and
Cauchy densities. The multivariate probability density function of
an image modelled by a $t$-distribution with mean
$\bmu_{l,d}(\alpha_{l,d}) =
\alpha_{l,d}\mathbf{G}_{d}\mathbf{s}_{l}$, scale $\delta_{l,d}$
and dof $\beta_{l,d}$ can be defined as

\begin{eqnarray}\label{pT}\nonumber
    \mathcal{T}(\mathbf{s}_{l}|\bmu_{l,d}, \delta_{l,d}, \beta_{l,d}) & = &
    \frac{\Gamma((N+\beta_{l,d})/2)}{\Gamma(\beta_{l,d}/2)(\pi\beta_{l,d}\delta_{l,d})^{N/2}}\\
    & & \times \left[1+\frac{||\mathbf{s}_{l} - \bmu_{l,d}||^2}{\beta_{l,d}\delta_{l,d}}\right]^{-(N+\beta_{l,d})/2}
\end{eqnarray}
where $\Gamma(.)$ is the Gamma function. Using a latent variable,
i.e. $\nu_{l,d}$, the $t$-distribution can also be written in
implicit form using a Gaussian and a Gamma density \citep{Liu95}:

\begin{equation}\nonumber
\mathcal{T}(\mathbf{s}_{l}|\bmu_{l,d}, \delta_{l,d}, \beta_{l,d})
=
\end{equation}
\begin{equation}\label{jointT}\nonumber
  \int
    \mathcal{N}\left(\mathbf{s}_{l}|\bmu_{l,d},\frac{\delta_{l,d}\mathbf{I}_{N}}{\nu_{l,d}}\right)
    \mathcal{G}\left(\nu_{l,d}|\frac{\beta_{l,d}}{2},\frac{\beta_{l,d}}{2}\right)d\nu_{l,d}.
\end{equation}

We use the representation in (\ref{jointT}) to estimate the
parameters using EM method.

We can write the density of $\mathbf{s}_{l}$ by using the image
differentials in different directions, by assuming directional
independence, as

\begin{equation}\label{priorsl}
        p(\mathbf{s}_{l}|\Theta)
        = \prod_{d=1}^{4}
        \mathcal{T}(\mathbf{s}_{l}|\bmu_{l,d}(\alpha_{l,d}), \delta_{l,d}, \beta_{l,d}).
\end{equation}
where $\Theta = \{ \alpha_{1:L,1:4}, \beta_{1:L,1:4},
\delta_{1:L,1:4} \}$ is the set of all parameter.

We assume uniform priors for $\alpha_{l,d}$ and $\delta_{l,d}$ and
use uninformative Jeffrey's prior for $\beta_{l,d}$; $\beta_{l,d}$
$\sim$ $1/\beta_{l,d}$.

\subsection{Likelihood}

Since the observation noise is assumed to be independent and
identically distributed zero-mean Gaussian at each pixel, the
likelihood is expressed as

\begin{eqnarray}\label{l}
    p(\mathbf{y}_{1:K} | \mathbf{s}_{1:L}, \mathbf{A}) &\propto& \prod_{k=1}^{K}  \exp  \left\{ - W(\mathbf{s}_{1:L}|\mathbf{y}_{k},\mathbf{A},\sigma_{k}^{2}) \right\} \\
    \label{l2} W(\mathbf{s}_{1:L}|\mathbf{y}_{k},\mathbf{A},\sigma_{k}^{2}) &=&
    \frac{||(\mathbf{y}_{k}-\mathbf{H}_{k}\sum_{l=1}^{L}a_{k,l}\mathbf{s}_{l})||^2}{2\sigma_{k}^{2}}
\end{eqnarray}
where $\mathbf{y}_{1:K}$ and $\mathbf{s}_{1:L}$ represent the set
of all observed and source images. The mixing matrix $\mathbf{A}$
contains all the mixing coefficients $a_{k,l}$ introduced in
(\ref{mix2}). We assume uniform priors for $a_{k,l}$. Matrix
$\mathbf{H}_{k}$ is the Toeplitz convolution matrix constituted by
$\mathbf{h}_{k}$ introduced in (\ref{mix2}).

\subsection{Posteriors}\label{sectionjopost}

By taking into account the parameters of the source priors, we
write the joint posterior density of all unknowns as:

\begin{equation}\label{bayesfull}
    p(\mathbf{s}_{1:L}, \mathbf{A},\Theta|\mathbf{y}_{1:K}) \propto
     p(\mathbf{y}_{1:K}|\mathbf{s}_{1:L}, \mathbf{A}) p(\mathbf{s}_{1:L}, \mathbf{A},\Theta)
\end{equation}
where $p(\mathbf{y}_{1:K} | \mathbf{s}_{1:L}, \mathbf{A})$ is the
likelihood and $p(\mathbf{s}_{1:L}, \mathbf{A},\Theta)$ is the
joint prior density of unknowns. The joint prior can be factorized
as $p(\mathbf{s}_{1:L}|\alpha_{1:L,1:4}, \beta_{1:L,1:4},
\delta_{1:L,1:4})$ $p(\mathbf{A})$ $p(\beta_{1:L,1:4})$
$p(\delta_{1:L,1:4})$ $p(\alpha_{1:L,1:4})$. Furthermore, since
the sources are assumed to be independent, the joint probability
density of the sources is also factorized as

\begin{equation}\label{ps}
  p(\mathbf{s}_{1:L}|\Theta)=\prod_{l=1}^{L}p(\mathbf{s}_{l}|\Theta)
\end{equation}

For estimating all of the unknowns, we write their conditional
posteriors as

\begin{eqnarray}\label{post}
\nonumber% \nonumber to remove numbering (before each equation)
  p(a_{k,l}|\mathbf{y}_{1:K},\mathbf{s}_{1:L}, \mathbf{A}_{-a_{k,l}},\Theta) & \propto  & p(\mathbf{y}_{1:K}|\mathbf{s}_{1:L},\mathbf{A})\\ \nonumber
  p(\alpha_{l,d}|\mathbf{y}_{1:K},\mathbf{s}_{1:L}, \mathbf{A},\Theta_{-\alpha_{l,d}}) & \propto & p(\mathbf{s}_{l}|\Theta) \\
  p(\beta_{l,d}|\mathbf{y}_{1:K},\mathbf{s}_{1:L}, \mathbf{A},\Theta_{-\beta_{l,d}}) & \propto & p(\mathbf{s}_{l}|\Theta)p(\beta_{l,d}) \\\nonumber
  p(\delta_{l,d}|\mathbf{y}_{1:K},\mathbf{s}_{1:L}, \mathbf{A},\Theta_{-\delta_{l,d}})  & \propto &  p(\mathbf{s}_{l}|\Theta)
  \\ \nonumber
   p(\mathbf{s}_{l}|\mathbf{y}_{1:K},\mathbf{s}_{(1:L)-l}, \mathbf{A},\Theta)  & \propto & p(\mathbf{y}_{1:K}|\mathbf{s}_{1:L},\mathbf{A}) p(\mathbf{s}_{l}|\Theta)
\end{eqnarray}
where "--variable" expressions in the subscripts denote the
removal of that variable from the variable set.

The ML estimation of the parameters $\alpha_{l,d}$, $\beta_{l,d}$
and $\delta_{l,d}$ using the EM method \citep{Liu95} is given in
Section \ref{BolReg}. To estimate the source images, we use a
version of the posterior $p(\mathbf{s}_{l}|.)$ augmented by
auxiliary variables and find the estimate by means of a Langevin
sampler. The details are given in Section \ref{algo}.

\section{Estimation of Astrophysical Maps and Parameters} \label{algo}

In this section, we give the estimation of the mixing matrix,
source maps and their parameters.

\subsection{Mixing Matrix} \label{BolMixMat}

We assume that the prior of $\mathbf{A}$ is uniform between 0 and
$\infty$. From (\ref{post}), it can be seen that the posterior
density of $a_{k,l}$ only depends on the Gaussian likelihood in
(\ref{l}). We can find the maximum likelihood estimate of
$a_{k,l}$ as

\begin{equation}\label{anupda}
    a_{k,l}  =  \frac{1}{\mathbf{s}_{l}^{T}\mathbf{H}_{k}^{T}\mathbf{H}_{k}\mathbf{s}_{l}} \mathbf{s}_{l}^{T}\mathbf{H}_{k}^{T}(\mathbf{y}_{k} - \mathbf{H}_{k} \sum_{i=1, i\neq
    l}^{L} a_{k,i}\mathbf{s}_{i}) u(a_{k,l})
\end{equation}
where $u(a_{k,l})$ is the unit step function.

\subsection{Astrophysical Map Estimation} \label{sectionmojopost}

We simulate the astrophysical maps from their posteriors using an
MCMC scheme. In the classical MCMC schemes, a random walk process
is used to produce the proposal samples. Although random walk is
simple, it affects the convergence time adversely. The random walk
process uses only the previous sample for producing a new
proposal. Instead of a random walk, we use the Langevin stochastic
equation, which exploits the gradient information of the energy
function to produce a new proposal. Since the gradient directs the
proposed samples towards the mode, the final sample set will
mostly come from around the mode of the posterior. The Langevin
equation used in this study is written as

\begin{equation}\label{lanvegin}
    \mathbf{s}_{l}^{k+1} = \mathbf{s}_{l}^{k} - \frac{1}{2}\mathbf{D}\mathbf{g}(\mathbf{s}_{1:L}^{k})
    + \mathbf{D}^{\frac{1}{2}} \mathbf{w}_{l}
\end{equation}
where $\mathbf{g}(\mathbf{s}_{1:L}^{k}) = [\nabla_{\mathbf{s}_{l}}
E(\mathbf{s}_{1:L})]_{\mathbf{s}_{1:L} = \mathbf{s}_{1:L}^{k}}$,
$\nabla_{\mathbf{s}_{l}}$ is the gradient with respect to
$\mathbf{s}_{l}$ and the diagonal matrix
$\mathbf{D}^{\frac{1}{2}}$ contains the discrete time steps
$\tau_{l,n}$, $n=1:N$. The total energy function
$E(\mathbf{s}_{1:L})$ is proportional to the negative logarithm of
the posterior as $-\log
p(\mathbf{s}_{l}|\mathbf{y}_{1:K},\mathbf{s}_{(1:L)-l},
\mathbf{A},\Theta)$. For the $i$th pixel, the diffusion
coefficient is $\mathbf{D}_{n,n}=\tau_{l,n}^{2}$. Here, matrix
$\mathbf{D}$ is referred to as the diffusion matrix. We determine it
by taking the inverse of the diagonal of the Hessian matrix of
$E(\mathbf{s}_{1:L})$. Rather than the expectation of the inverse
of Hessian matrix, we use its diagonal calculated by the value of
$\mathbf{s}_{l}$ at the discrete time $k$ as
\citep{Becker89,Kayabol10}

\begin{equation}\label{Difpar0}
    \mathbf{D} = 2 [\langle
    \overline{\mathcal{H}}(\mathbf{s}_{l}^{k})\rangle]^{-1}.
\end{equation}
where $\overline{\mathcal{H}}(\mathbf{s}_{l}^{k}) = \mathrm{diag}
\left\{\mathcal{H}(\mathbf{s}_{l})
\right\}_{\mathbf{s}_{l}=\mathbf{s}_{l}^{k}}$ and the operator
$\mathrm{diag}\{.\}$ extracts the main diagonal of the Hessian
matrix.

Since the random variables for the image pixel intensities are
produced in parallel by using (\ref{lanvegin}), the procedure is
faster than the random walk process adopted in
\citep{Kayabol09ip}. Equation (\ref{lanvegin}) produces a
candidate map sample by taking into account the noise, the
channel-variant blur and the mixing matrix. Unlike LS solution,
this equation does not contain any matrix inversion. The
derivation details of the equation can be found in
\citep{Kayabol10}.

After the sample production process, the samples are applied to a
Metropolis-Hastings \citep{Hastings} scheme pixel-by-pixel. The
acceptance probability of any proposed sample is defined as
$\min\{\varphi(s_{l,n}^{k+1},s_{l,n}^{k}),1\}$, where

\begin{equation}\label{acptprob}
    \varphi(s_{l,n}^{k+1},s_{l,n}^{k}) \propto
     e^{ - \Delta E(s_{l,n}^{k+1})}\frac{q(s_{l,n}^{k}|s_{l,n}^{k+1})}{q(s_{l,n}^{k+1}|s_{l,n}^{k})}
\end{equation}
where $\Delta E(s_{l,n}^{k+1}) = E(s_{l,n}^{k+1}) -
E(s_{l,n}^{k})$ and $E(s_{l,n}^{k}) = W(s_{1:L,n}^{k}) +
U(s_{l,n}^{k})$. For any single pixel, $U(s_{l,n})$ can be derived
from (\ref{pT}) and (\ref{priorsl}) as

\begin{equation}\label{Using}
    U(s_{l,n}) = \sum_{d=1}^{D} \frac{1+\beta_{l,d}}{2}\log\left[1+\frac{\phi_{d}(s_{l,n},\alpha_{l,d})}{\beta_{l,d}\delta_{l,d}}\right]
\end{equation}

The proposal density $q(s_{l,n}^{k+1}|s_{l,n}^{k})$ is obtained,
from (\ref{lanvegin}), as

\begin{equation}\label{q}
    \mathcal{N} \left(s_{l,n}^{k+1}|s_{l,n}^{k} +
\frac{\tau_{l,n}}{2} g(s_{l,n}^{k}), \tau_{l,n}^{2} \right)
\end{equation}

The Metropolis-Hastings steps and the Langevin proposal equation are
embedded into the main algorithm, as detailed in Appendix
\ref{ALS}. With this algorithm, we approach the solution
iteratively, avoiding the inversion of the convolution matrix
$\mathbf{H}_{k}$ and the mixing matrix $\mathbf{A}$.

\subsection{Parameters of $t$-distribution } \label{BolReg}

We find the mode estimates of the parameters of the
$t$-distribution using EM method. We can write the joint posterior
of the parameters $\alpha_{l,d}$, $\beta_{l,d}$ and $\delta_{l,d}$
such that
$p(\alpha_{l,d},\beta_{l,d},\delta_{l,d}|\mathbf{t}_{l,d},\Theta_{-\{\alpha_{l,d},\beta_{l,d},\delta_{l,d}\}})
= p(\mathbf{t}_{l,d}|\Theta)p(\beta_{l,d})$. In EM, rather than
maximizing $\log
\left\{p(\mathbf{t}_{l,d}|\Theta)p(\beta_{l,d})\right\}$, we
maximize the following function iteratively

\begin{equation}\label{maxem}
    \Theta^{k+1} = \arg \max_{\Theta} Q(\Theta;\Theta^{k})
\end{equation}
where superscript $k$ represents the iteration number and

\begin{equation}\label{emyA}
  Q(\Theta;\Theta^{k}) =  \left \langle \log
  \{p(\mathbf{t}_{l,d}|\Theta)p(\beta_{l,d}))\}\right\rangle_{\nu_{l,d}|\mathbf{t}_{l,d}^{k},\Theta^{k}}
\end{equation}
where $p(\nu_{l,d}|\mathbf{t}_{l,d}^{k},\Theta^{k})$ is the
posterior density of the hidden variable $\nu_{l,d}$ conditioned
on parameters estimated in the previous step $k$ and $\left\langle
. \right\rangle_{\nu_{l,d}|\mathbf{t}_{l,d}^{k},\Theta^{k}}$
represents the expectation with respect to
$\nu_{l,d}|\mathbf{t}_{l,d}^{k},\Theta^{k}$. For simplicity,
hereafter we use only the notation $\left\langle . \right\rangle$
to represent this expectation.

In the E (expectation) step of the EM algorithm, the posterior
expectation of $\nu_{l,d}$ is found as in \cite{Kayabol10}

\begin{equation}\label{expectnu}
    \langle\nu_{l,d}\rangle
    = \frac{N+\beta_{l,d}^{k}}{\beta_{l,d}^{k}} \left(1+\frac{\phi_{d}(\mathbf{s}_{l}^{k},\alpha_{l,d}^{k})}{\beta_{l,d}^{k}\delta_{l,d}^{k}}\right)^{-1}
\end{equation}

In the M (maximization) step, (\ref{emyA}) is maximized with
respect to $\Theta$. To maximize this function, we alternate among
the variables $\alpha_{l,d}$, $\beta_{l,d}$ and $\delta_{l,d}$.
The solutions are found as

\begin{equation}\label{alphaML}
    \alpha_{l,d} = \frac{\mathbf{s}_{l}^{T}\mathbf{G}_{d}^{T}\mathbf{s}_{l}}{\mathbf{s}_{l}^{T}\mathbf{G}_{d}^{T}\mathbf{G}_{d}\mathbf{s}_{l}}
\end{equation}

\begin{equation}\label{deltaML}
    \delta_{l,d} = \langle\nu_{l,d}\rangle\frac{\phi_{d}(\mathbf{s}_{l},\alpha_{l,d})}{N}
\end{equation}

The maximization with respect to $\beta_{l,d}$ does not have a
simple solution. It can be solved by setting its first derivative
to zero:

\begin{equation}\label{betazero}
    - \psi_{1}(\beta_{l,d}/2) + \log\beta_{l,d} + \langle\log\nu_{l,d}\rangle - \langle\nu_{l,d}\rangle +1 = 0
\end{equation}
where $\psi_{1}(.)$ is the first derivative of $\log\Gamma(.)$ and
it is called digamma function.

\subsection{Technical Details of the Algorithm} \label{BolAlg}

We represent the proposed Adaptive Langevin Sampler (ALS)
algorithm in Appendix \ref{ALS}. The symbol $\longleftarrow$
denotes analytical update, the symbol $\longleftarrow_{0}$ denotes
update by finding zero and the symbol $\sim$ denotes the update by
random sampling. The sampling of the sources is done by the
Metropolis-Hastings scheme with Langevin proposal equation. The
random map produced by Langevin proposal is applied to a threshold
function to keep the intensities of the maps in the physical
margins. We have used the following margins for CMB, synchrotron,
dust and free-free, respectively, $[-0.45,0.45]$, $[0,0.5]$,
$[0,25]$ and $[0,0.1]$. They are in antenna temperature $(\Delta
T)_{A}$ in mK same as the maps. We have determined these margins
by using five patches from our simulations.

\subsubsection{Initialization} \label{ini}

By considering  previous studies  \citep{Bonaldi07, Ricciardi10},
we assume that the parametric mixing matrix is known with an
error. We initialize the mixing matrix by using the spectral
indices obtained in \citep{Bonaldi07, Ricciardi10}. The parametric
model is formed so that the columns of synchrotron and dust vary
according to power laws only depending on one spectral index.
Previous experiments show that the error in spectral index of
synchrotron changes from patch to patch and takes a maximum value
of about 1.72$\%$. For spectral index of dust, the maximum error
is about 0.58$\%$. We have fixed the columns of CMB and free-free
as they are known. We obtain realistic observations by mixing
components with a mixing matrix which is formed by using the
spectral indices 2.9 for synchrotron and 1.8 for dust. In the
reconstruction part, we assume that the spectral indices are
estimated with an error of 1.72$\%$ for synchrotron and 0.58$\%$
for dust. So, we initialize the mixing matrix values to maximum
error case such that the spectral indices are equal to 2.85 for
synchrotron and 1.7894 for dust.

To estimate the spectral indices, one can use the FDCCA (Fourier
Domain Correlated Component Analysis) \citep{Bedini07} method, but
it is not necessary to use this algorithm. Non-parametric mixing
matrix estimation methods can be used, such as Independent
Component Analysis (ICA) \citep{Hyvarinen97} or Spectral Matching
ICA (SMICA) \citep{Cardoso02}.

To initialize the component maps, we ignore the antenna beams and
apply the inverse of the initial mixing matrix to the raw
observations directly. If we denote the initial mixing matrix
$\mathbf{A}^{0}$, we initialize the maps with LS solution as
$\mathbf{s}^{0}(n)=((\mathbf{A}^{0})^{T}\mathbf{A}^{0})^{-1}(\mathbf{A}^{0})^{T}\mathbf{y}(n)$
where the vector $\mathbf{y}(n)$ contains the observation
intensities at $n^{th}$ pixel. LS solution is not a good solution
since it does not take the noise and the resolutions of the
observations into consideration, but it provides a simple solution
without any preprocessing intervention. In this way, our algorithm
starts with initial maps which are some linear combination of the
raw observations. The initial values of $\alpha_{l,d}$ can be
calculated directly from image differentials. We initialized the
$\beta_{l,d}^{0}=20$ and found the initial value of $\delta_{l,d}$
by equaling the expectation in (\ref{deltaML}) to a constant, in
this study, we take it to be equal to $1.5$. The initial value is
found as $\delta_{l,d}^{0} =
1.5\phi_{d}(\mathbf{s}_{l}^{0},\alpha_{l,d}^{0})/N$.

\subsubsection{Stopping Criterion}

We observe the normalized absolute difference
$\epsilon_{l}^{k}=|\mathbf{s}_{l}^{k}-\mathbf{s}_{l}^{k-1}|/|\mathbf{s}_{l}^{k-1}|$
between sequential values of $\mathbf{s}_{l}$ to decide the
convergence of the Markov Chain to an equilibrium. If the average
$\bar{\epsilon}_{l}^{k} =
\frac{1}{k}\sum_{t=1}^{k}\epsilon_{l}^{t} \leq 5e-2$, we assume
that the chain has converged to the equilibrium for
$\mathbf{s}_{l}$ and denote this point $T_{l}=k$. Since we have
$L$ parallel chains for $L$ sources, the ending point of the
burn-in period of the whole Monte Carlo chain is $T_{s} = \max_{l}
T_{l}$. We ignore the samples before $T_{s}$. We keep the
iteration going until $T_{e}$ that is the ending point of the post
burn-in period simulation. In the experiments, we have used 100
iterations after burn-in period, so $T_{e}=T_{s}+100$.

At the end of the simulation the final estimates of the component
maps are calculated as

\begin{equation}\label{mu}
    \hat{s}_{l,n} = \frac{1}{T_{e}-T_{s}}\sum _{k=T_{s}}^{T_{e}}
    s_{l,n}^{k}
\end{equation}

\section{Simulation Results} \label{simres}

In order to test our ideas, we have used a set of realistic
simulations obtained from the Planck Sky Model (PSM), a set of
maps and tools developed by the Planck Working Group 2 (WG2) team
as a fundamental part of the preparation for the Planck mission
\citep{Tauber10}. Apart from the CMB itself, the PSM contains
state-of-the-art simulations of all the relevant Galactic and
extragalactic astrophysical components; for this work we use a
simplified set of simulations that contains CMB and Galactic
(synchrotron, free-free and dust) components only, plus
instrumental noise. We have used simulations of the nine Planck
frequencies. The main characteristics of the simulations are
listed in Table \ref{QObserv}.

We have tested our algorithm on five different 128x128 patches
distributed along the central galactic meridian, and centered at
galactic coordinates (00,00), (00,20), (00,40), (00,60) and
(00,80). The actual size of the patches on the sky is
14.65$^{\circ}$ and the pixel size is 6.87 arcmin. The maps are in
antenna temperature $(\Delta T)_{A}$ in mK. The related noise
levels are presented in Table \ref{QObserv}. We model the blurring
functions as Gaussian shaped functions according to antenna
apertures. Their standard deviations in pixels are given in Table
\ref{QObserv}.

\begin{table*}
 \centering
 \begin{minipage}{140mm}
  \caption{Channel frequencies, the standard deviations of the related Gaussian point spread functions and the noise standard deviations in $(\Delta T)_{A}$ [mK].}
  \begin{tabular}{@{}lrrrrrrrrr@{}}
 \hline
 Channel\\ frequencies [GHz]  & 30 & 44 & 70 & 100 & 143 & 217 & 353 & 545 & 857 \\
 psf std\\ .[pixels] & 7.0069 & 4.8836 & 2.9726 & 2.1233 & 1.5075 & 1.0617 & 1.0617 & 1.0617 & 1.0617\\
 Noise std\\ $(\Delta T)_{A}$ [mK] & 0.0259 & 0.0248 & 0.0233 & 0.0074 & 0.0038 & 0.0032 & 0.0023 & 0.0019 & 0.0009\\
\hline
\end{tabular}\label{QObserv}
\end{minipage}
\end{table*}

\begin{figure*} \centering
  \includegraphics[width=6.6in]{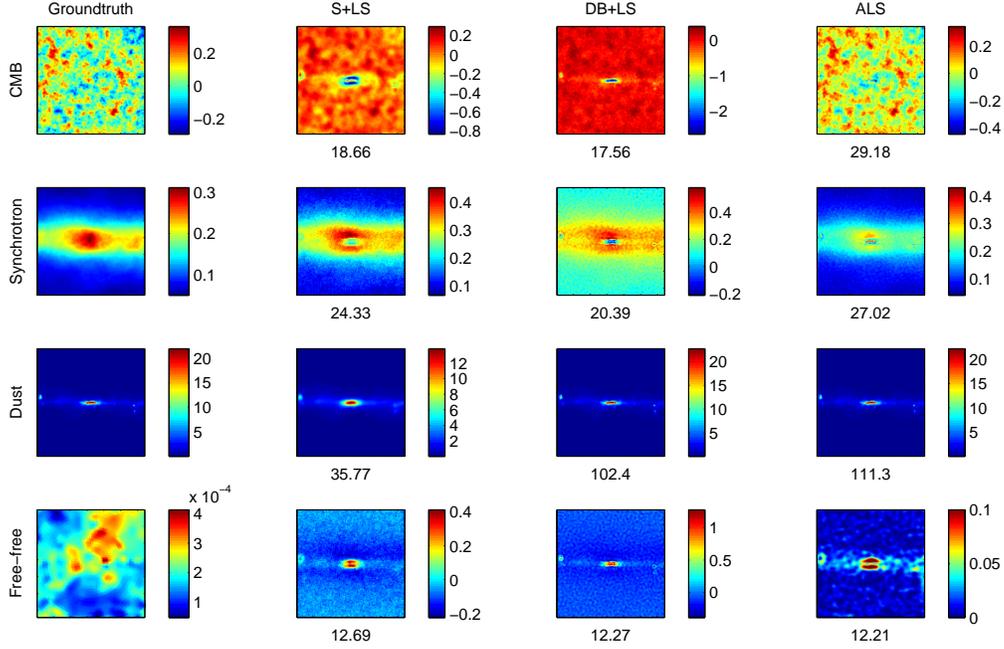}
\caption{The estimated astrophysical maps at 100 GHz reference
frequency from blurred and noisy observations with the S+LS, DB+LS
and proposed ALS methods. The location of the patch is $0^{\circ}$
galactic longitude and  $0^{\circ}$ latitude. The $PSIR_{pix}$
values are denoted under the each map.}\label{Es0000}
\end{figure*}

\begin{figure*} \centering
  \includegraphics[width=6.6in]{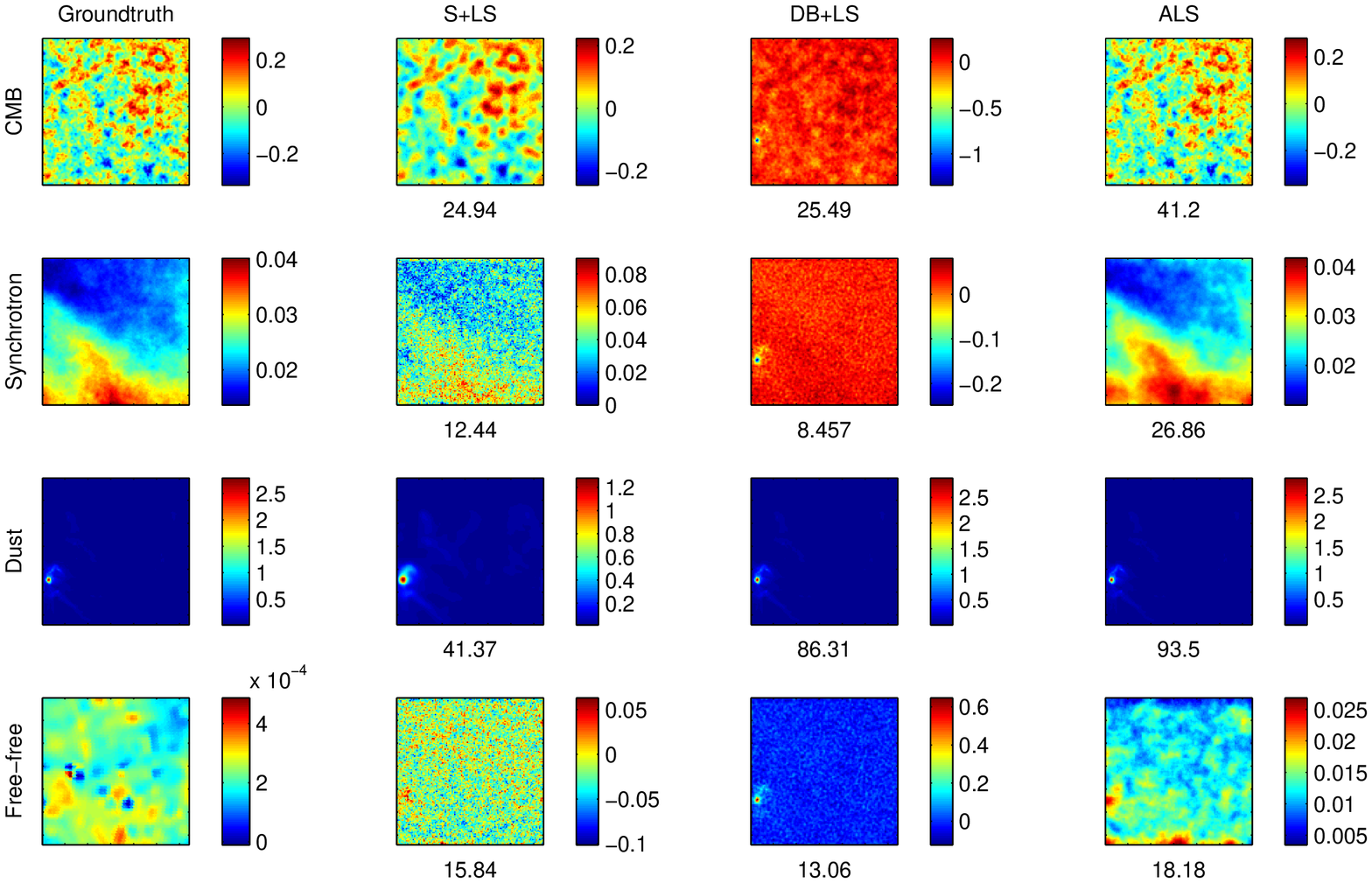}
  \caption{The estimated astrophysical maps at 100 GHz reference
frequency from blurred and noisy observations with the S+LS, DB+LS
and proposed ALS methods. The location of the patch is $0^{\circ}$
galactic longitude and  $20^{\circ}$ latitude. The $PSIR_{pix}$
values are denoted under the each map. }\label{Es0020}
\end{figure*}

\begin{figure*} \centering
  \includegraphics[width=6.6in]{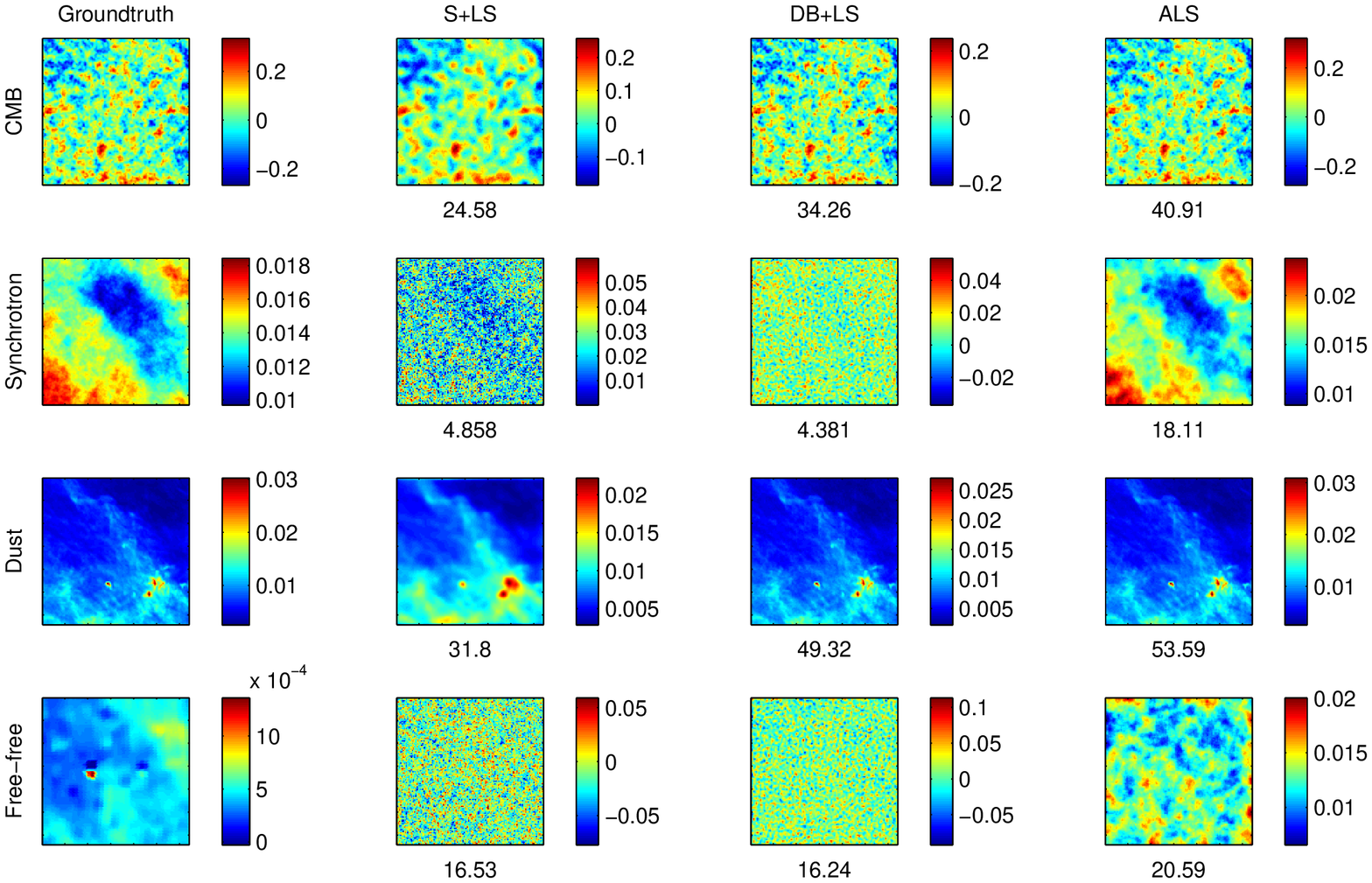}
  \caption{The estimated astrophysical maps at 100 GHz reference
frequency from blurred and noisy observations with the S+LS, DB+LS
and proposed ALS methods. The location of the patch is $0^{\circ}$
galactic longitude and  $40^{\circ}$ latitude. The $PSIR_{pix}$
values are denoted under the each map. }\label{Es0040}
\end{figure*}

\begin{figure*} \centering
  \includegraphics[width=6.6in]{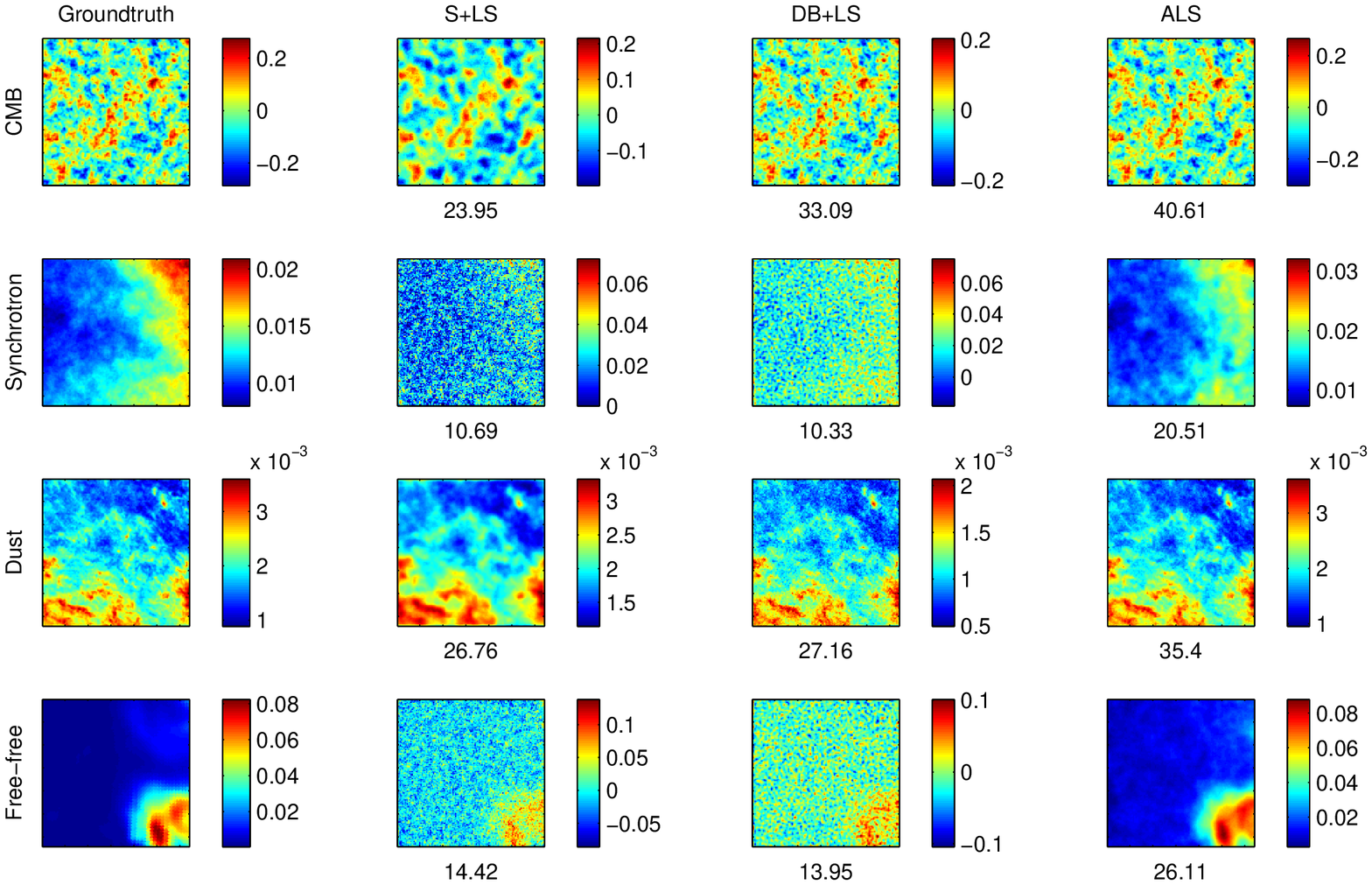}
  \caption{The estimated astrophysical maps at 100 GHz reference
frequency from blurred and noisy observations with the S+LS, DB+LS
and proposed ALS methods. The location of the patch is $0^{\circ}$
galactic longitude and  $60^{\circ}$ latitude. The $PSIR_{pix}$
values are denoted under the each map. }\label{Es0060}
\end{figure*}

\begin{figure*} \centering
  \includegraphics[width=6.6in]{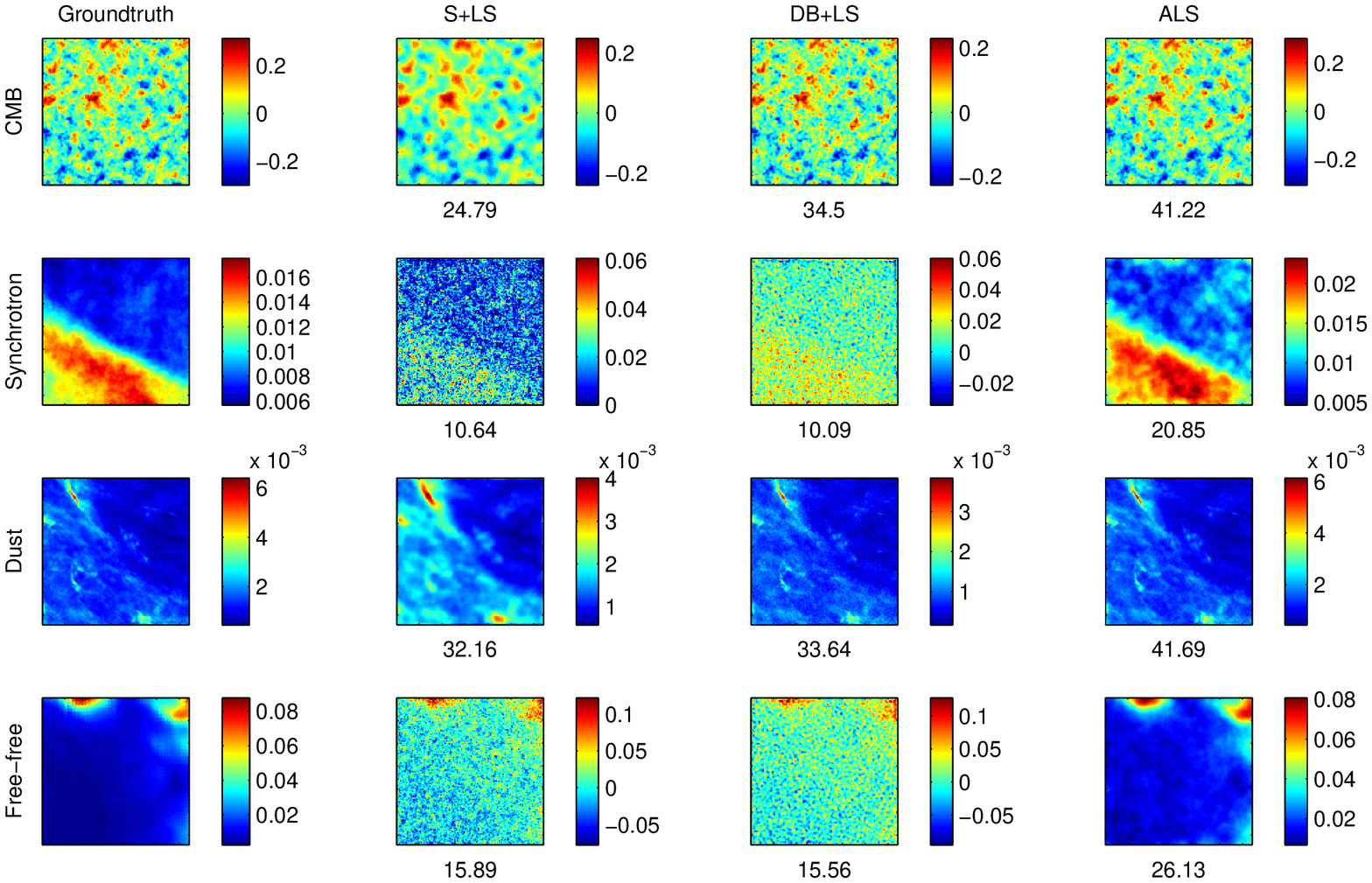}
  \caption{The estimated astrophysical maps at 100 GHz reference
frequency from blurred and noisy observations with the S+LS, DB+LS
and proposed ALS methods. The location of the patch is $0^{\circ}$
galactic longitude and  $80^{\circ}$ latitude. The $PSIR_{pix}$
values are denoted under the each map. }\label{Es0080}
\end{figure*}

We use three different performance measures defined in the pixel
domain to evaluate the success of the proposed algorithm among the
others. The Peak Signal-to-Interference Ratio ($PSIR_{pix}$) in
the pixel domain is defined as

\begin{equation}\label{psnr}
    PSIR_{pix} = 20\log\left(\frac{\sqrt{N}\max(\mathbf{s}^{\ast}_{l})}{||RE||}\right)
\end{equation}
where $RE = \mathbf{s}^{\ast}_{l}-\hat{\mathbf{s}}_{l}$ is the
Reconstruction Error between the ground-truth
$\mathbf{s}^{\ast}_{l}$ and the estimated image
$\hat{\mathbf{s}}_{l}$. We use this error measure instead of the
absolute difference between the ground-truth and the estimate,
because we need a normalized error to compare the error in large
variations of the intensities of the different components. The
logarithm gives a good observation possibility for the values
varying in a large scale. Fig. \ref{Es0000}, \ref{Es0020},
\ref{Es0040}, \ref{Es0060} and \ref{Es0080} show the estimated
maps located at the coordinates of (0,0), (0,20), (0,40), (0,60)
and (0,80).

We compare our proposed method ALS with the S+LS and DB+LS
solutions. S+LS solution is obtained by smoothing the observed
data to the resolution of the lowest resolution channel and then
applying the inverse of the mixing matrix to the equal resolution
maps. In DB+LS, we first apply a de-blurring (DB) process to
observation channels. For deblurring, we apply Wiener filter
separately to each channel using the known psf and the noise
level. We find the DB+LS solution by applying the inverse of the
mixing matrix to the deblurred maps.

For the patch (0,0), we obtained a good reconstruction for CMB
with the proposed method (Fig. \ref{Es0000}). In the middle of the
map, the effect of the dust has been already seen. The effect of
the dust also exists in the synchrotron map. The free-free
component radiation map in the patches (0,0), (0,20) and (0,40)
cannot be estimated by any method, since its intensity is very
weak. For all the patches, the proposed method reconstructs better
the CMB and the foreground maps in the sense of $PSIR_{pix}$.

We also estimate the error in the maps. Using MC samples, we can
find the uncertainty in the estimation. We call this MC error and
calculate its standard deviation for single pixel as

\begin{equation}\label{nee}
    \sigma_{MC} = \left( \frac{1}{T_{e}-T_{s}}\sum _{k=T_{s}}^{T_{e}}
    (\hat{s}_{l,n}-s_{l,n}^{k})^{2} \right)^{\frac{1}{2}}
\end{equation}
where the number $T_{e}$ is the ending point of the simulation. We
use 100 iterations after convergence, so in this study
$T_{e}=T_{s}+100$.

We obtain another error measure by fitting a Gaussian to the
posterior of the source image using the Laplace Approximation (LA)
method and calculating the standard deviation, $\sigma_{LA}$, of
the approximated Gaussian. Table \ref{sigmaer} lists the average
standard deviations $\overline{\sigma}_{RE}$,
$\overline{\sigma}_{MC}$ and $\overline{\sigma}_{LA}$ of the
reconstruction, the Monte Carlo and the Laplace Approximation (LA)
errors, respectively. The reconstruction error is always greater
than the estimated errors. MC and LA errors are quite close to
each other. The minimum errors for CMB are found in the patches
(0,20) and (0,40).
\begin{table}
  \caption{Average standard deviations of Reconstruction Error (RE), Monte Carlo (MC) uncertainty and Laplace Approximation (LA) uncertainty.}
  \label{sigmaer}
  \centering
\begin{tabular}{|c||c|c|c|c|}
  %\hline
  \multicolumn{1}{}{} &   \multicolumn{3}{c}{$(0^{\circ},0^{\circ})$}     \\
   \hline
   & CMB \hspace{8pt}& Synchrotron & Dust\hspace{7pt} & Free-Free\hspace{2pt}  \\
  \hline
   $\overline{\sigma}_{RE}$               & 30.49e-3 & 13.82e-3 & 15.04e-3 & 16.32e-3  \\
   $\overline{\sigma}_{MC}$               &  2.38e-3 &  0.42e-3 &  0.08e-3 &  1.50e-3  \\
   $\overline{\sigma}_{LA}$               &  3.38e-3 &  0.14e-3 &  0.06e-3 &  1.28e-3  \\
  \hline
\end{tabular}

\begin{tabular}{|c||c|c|c|c|}
  %\hline
  \multicolumn{1}{}{} &   \multicolumn{3}{c}{$(0^{\circ},20^{\circ})$}     \\
   \hline
   & CMB \hspace{8pt}& Synchrotron & Dust\hspace{7pt} & Free-Free\hspace{2pt}  \\
  \hline
   $\overline{\sigma}_{RE}$                & 14.83e-3 & 2.26e-3 & 1.03e-3 & 12.80e-3  \\
   $\overline{\sigma}_{MC}$               &  2.81e-3 & 1.04e-3 & 0.03e-3 &  2.10e-3  \\
   $\overline{\sigma}_{LA}$               &  3.37e-3 & 0.92e-3 & 0.06e-3 &  1.46e-3  \\
  \hline
\end{tabular}

\begin{tabular}{|c||c|c|c|c|}
  %\hline
  \multicolumn{1}{}{} &   \multicolumn{3}{c}{$(0^{\circ},40^{\circ})$}     \\
   \hline
   & CMB \hspace{8pt}& Synchrotron & Dust\hspace{7pt} & Free-Free\hspace{2pt}  \\
  \hline
   $\overline{\sigma}_{RE}$                & 14.22e-3 & 2.47e-3 & 0.17e-3 & 12.25e-3  \\
   $\overline{\sigma}_{MC}$               &  2.88e-3 & 1.28e-3 & 0.03e-3 &  1.96e-3  \\
   $\overline{\sigma}_{LA}$               &  3.37e-3 & 1.12e-3 & 0.02e-3 &  1.48e-3  \\
  \hline
\end{tabular}

\begin{tabular}{|c||c|c|c|c|}
  %\hline
  \multicolumn{1}{}{} &   \multicolumn{3}{c}{$(0^{\circ},60^{\circ})$}     \\
   \hline
   & CMB \hspace{8pt}& Synchrotron & Dust\hspace{7pt} & Free-Free\hspace{2pt}  \\
  \hline
   $\overline{\sigma}_{RE}$                & 10.72e-3 & 2.85e-3 & 0.06e-3 & 8.27e-3  \\
   $\overline{\sigma}_{MC}$               &  2.72e-3 & 1.34e-3 & 0.02e-3 & 2.05e-3  \\
   $\overline{\sigma}_{LA}$               &  3.37e-3 & 1.05e-3 & 0.02e-3 & 1.53e-3  \\
  \hline
\end{tabular}

\begin{tabular}{|c||c|c|c|c|}
  %\hline
  \multicolumn{1}{}{} &   \multicolumn{3}{c}{$(0^{\circ},80^{\circ})$}     \\
   \hline
   & CMB \hspace{8pt}& Synchrotron & Dust\hspace{7pt} & Free-Free\hspace{2pt}  \\
  \hline
   $\overline{\sigma}_{RE}$                & 10.29e-3 & 3.21e-3 & 0.06e-3 & 7.40e-3  \\
   $\overline{\sigma}_{MC}$               &  2.83e-3 & 1.43e-3 & 0.02e-3 & 2.11e-3  \\
   $\overline{\sigma}_{LA}$               &  3.37e-3 & 1.11e-3 & 0.02e-3 & 1.54e-3  \\
  \hline
\end{tabular}
\end{table}

\begin{figure*}%
\centering \subfigure[Patch
$(0^{\circ},0^{\circ})$]{\includegraphics[width=3.4in]{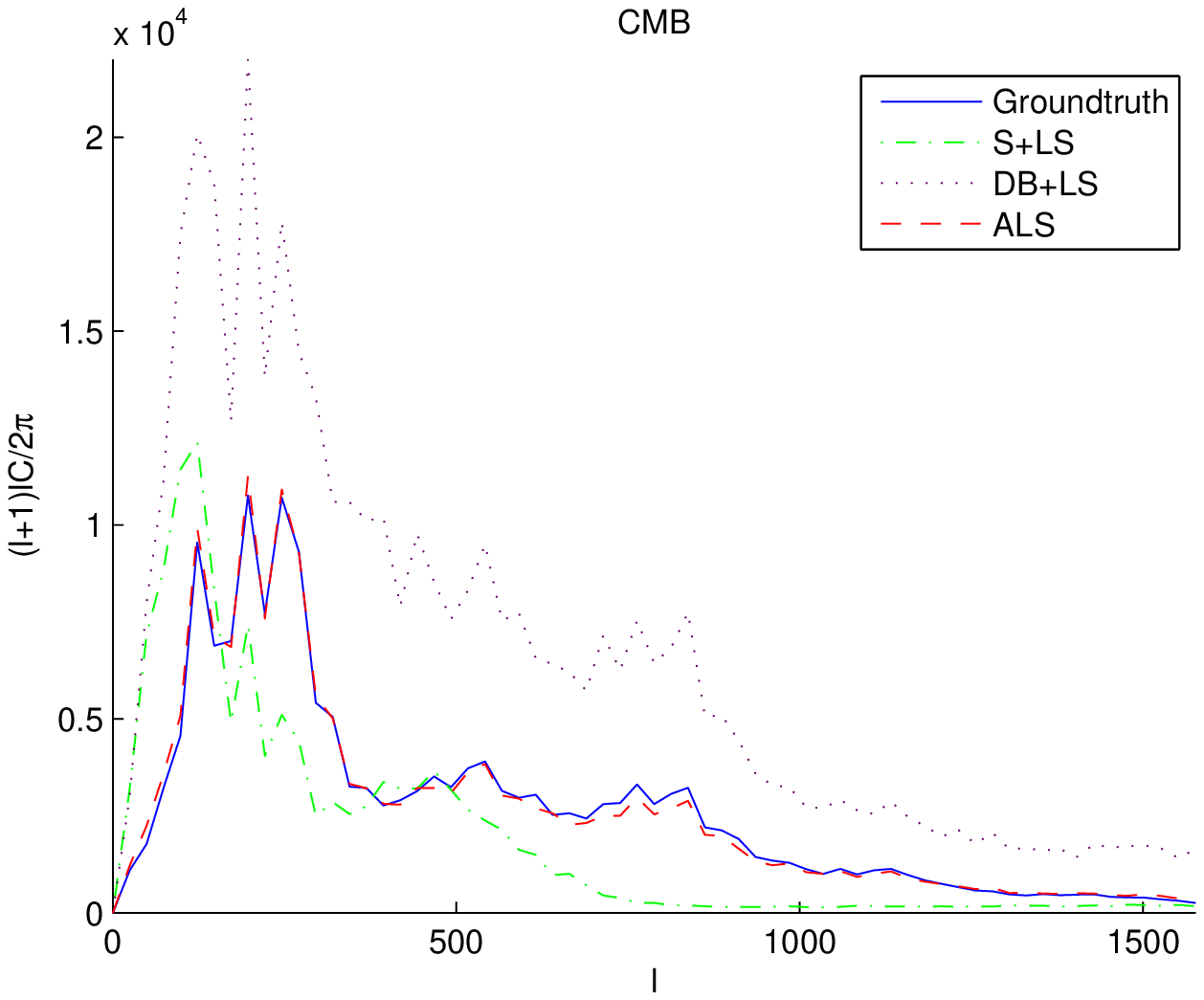}
\label{fig_first_case}} \hfil \subfigure[Patch
$(0^{\circ},20^{\circ})$]{\includegraphics[width=3.4in]{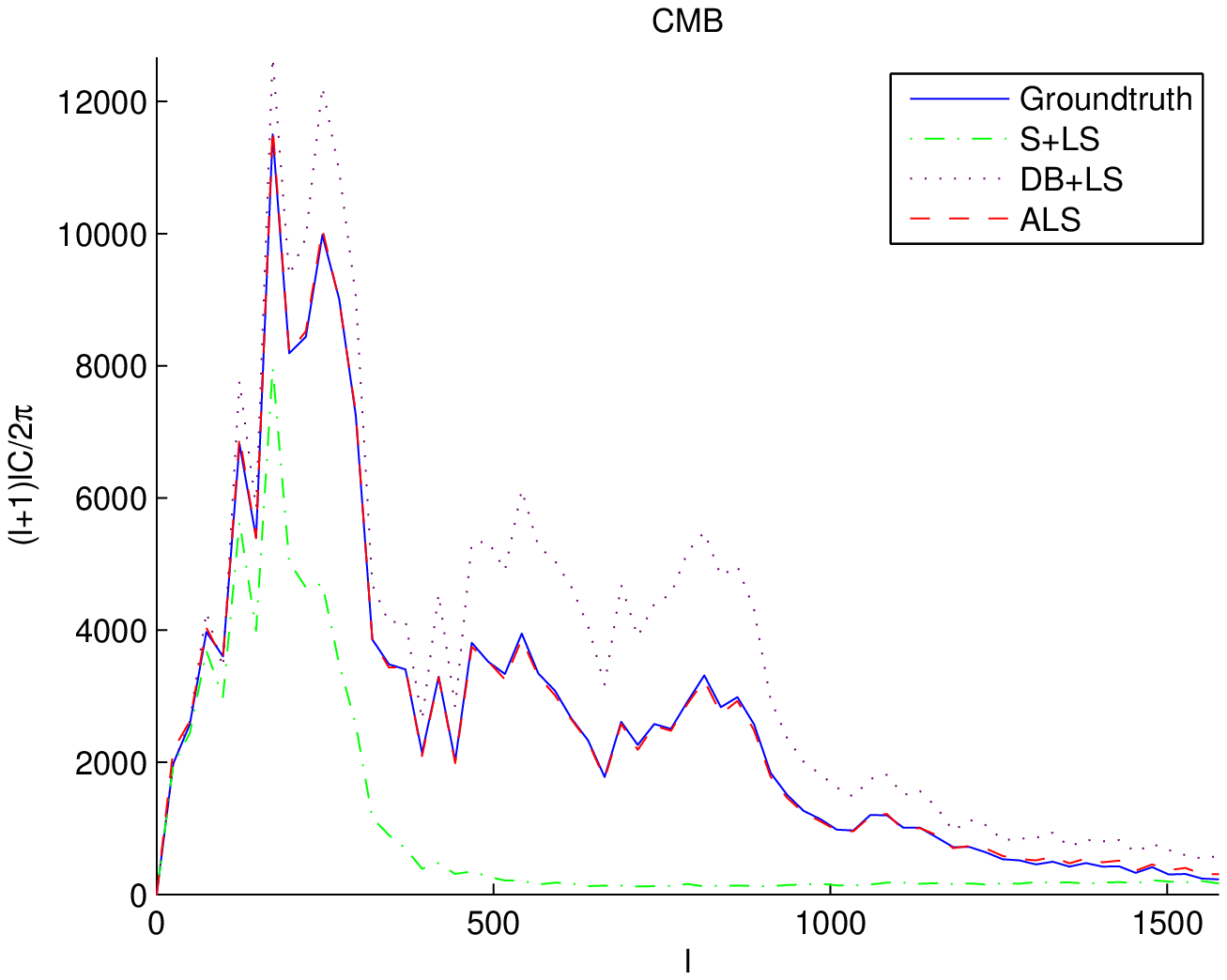}
\label{fig_first_case}}\\

\subfigure[Patch
$(0^{\circ},40^{\circ})$]{\includegraphics[width=3.4in]{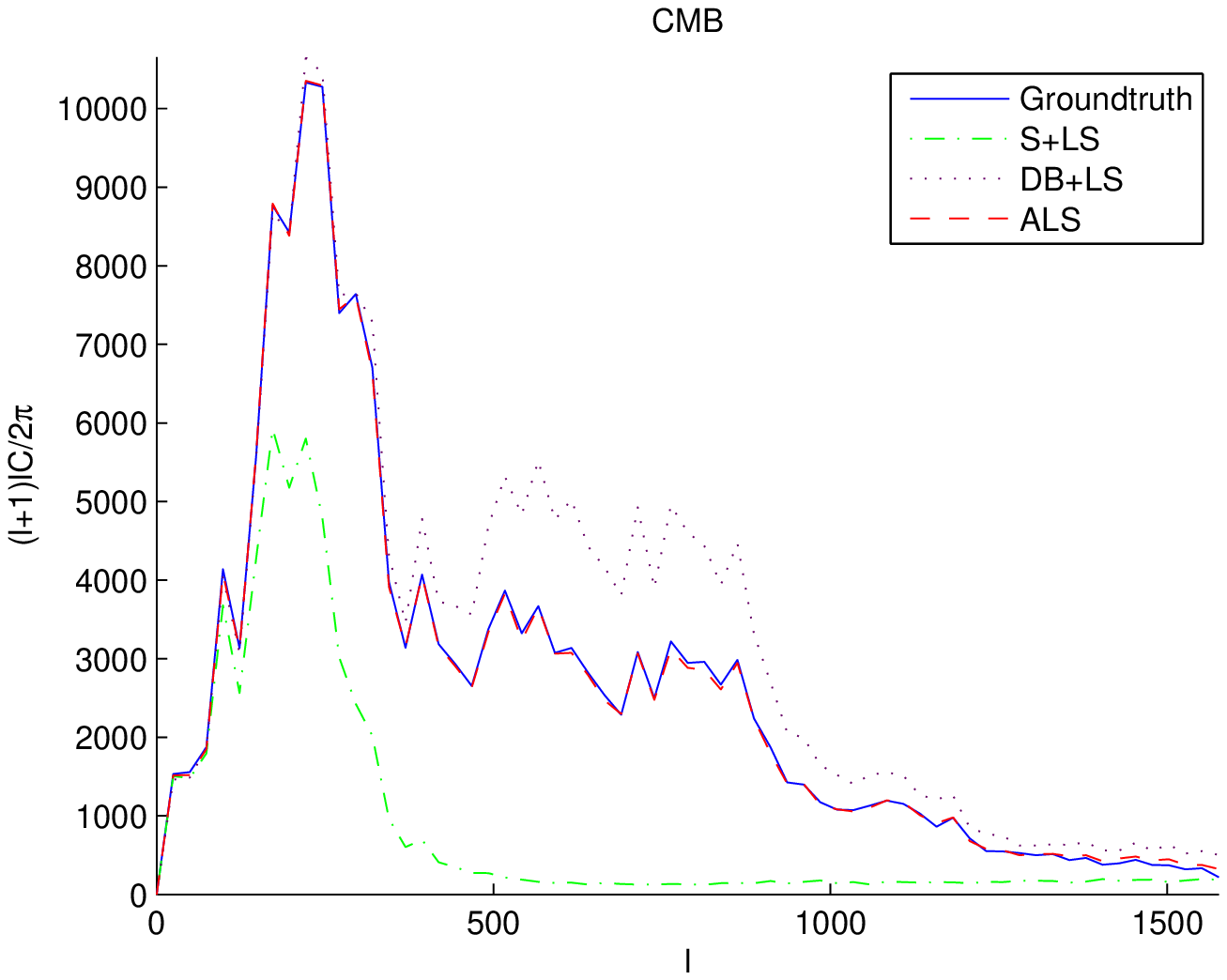}
\label{fig_second_case}} \hfil \subfigure[Patch
$(0^{\circ},60^{\circ})$]{\includegraphics[width=3.4in]{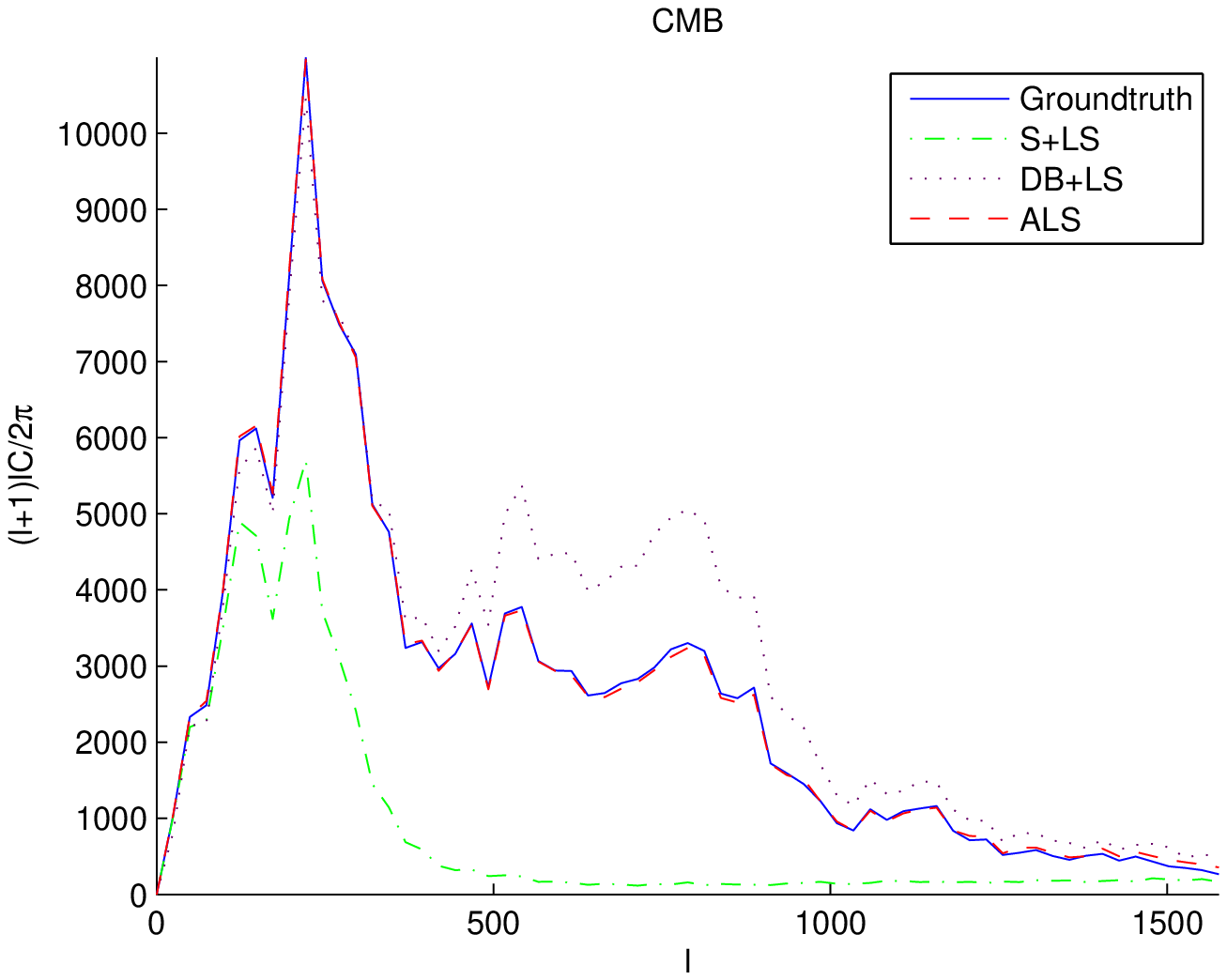}
\label{fig_second_case}}\\
\subfigure[Patch
$(0^{\circ},80^{\circ})$]{\includegraphics[width=3.4in]{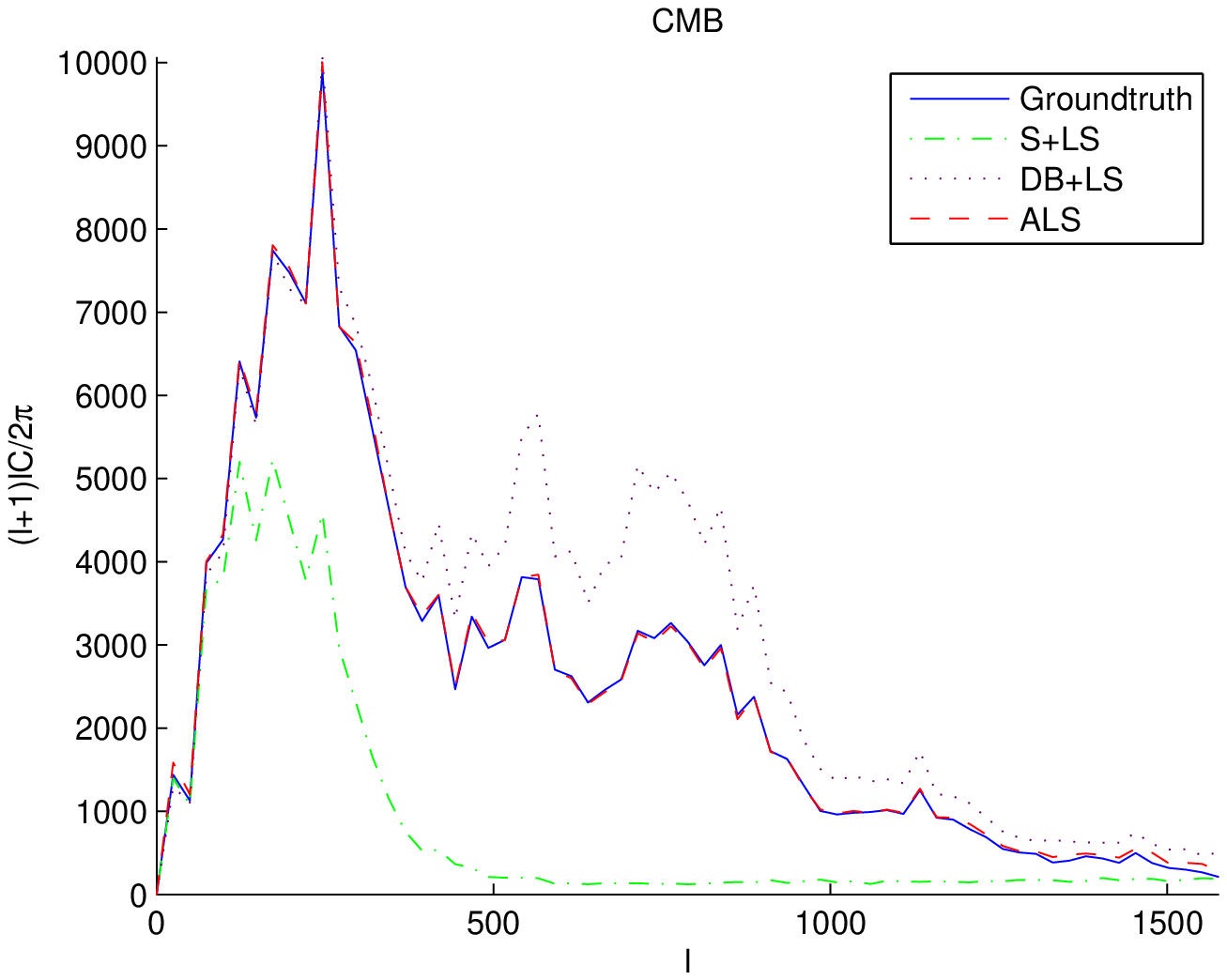}
\label{fig_second_case}}%
 \caption{Comparison of the standard power spectrum of the ground-truth maps located at a) $(0^{\circ},0^{\circ})$, b) $(0^{\circ},20^{\circ})$, c)
  $(0^{\circ},40^{\circ})$, d) $(0^{\circ},60^{\circ})$ and e) $(0^{\circ},80^{\circ})$ with those obtained by S+LS, DB+LS and proposed ALS methods.} \label{freq}
\end{figure*}

The plots in Fig. \ref{freq} compare the angular power spectrum,
$\overline{C}_{\ell}$ defined as $\overline{C}_{\ell} =
(\ell+1)\ell C_{\ell}/2\pi$, of the ground-truth CMB maps and
those obtained by S+LS, DB+LS and the proposed ALS methods. To
plot $\overline{C}_{\ell}$, we sample it by taking $\sqrt{N}/2+1$
samples in the interval $[0,\ell_{\max}]$ where
$\ell_{\max}=180/14.65\sqrt{N}$. All the patches, the power
spectra found by the proposed ALS method fit the ground-truth
spectra more tightly than the others. The S+LS method gives bad
results in the high frequency regions of the spectrums because of
smoothing. The DB+LS method causes an attenuation in the low
frequency regions. The root mean square error between the
groundtruth angular power spectrum of CMB and those obtained by
S+LS, DB+LS and proposed ALS methods are presented in Table
\ref{rmsCMB}. The proposed method provide one order of the
magnitude better fit than the others.

\begin{table}
\caption{Root mean square error between the groundtruth standard
power spectrum of CMB and those obtained by S+LS, DB+LS and
proposed ALS methods at patches $(0^{\circ},0^{\circ})$,
$(0^{\circ},20^{\circ})$, $(0^{\circ},40^{\circ})$,
$(0^{\circ},60^{\circ})$ and $(0^{\circ},80^{\circ})$.}
  \label{rmsCMB}
  \centering
\begin{tabular}{|c||c|c|c|}
  %\hline
  %\multicolumn{1}{}{} &   \multicolumn{3}{c}{0,0}     \\
   \hline
   & S+LS & DB+LS & ALS  \\
  \hline
   $(0^{\circ},0^{\circ})$      & 2.1798e+3 & 7.4638e+3 &  0.1894e+3 \\
   $(0^{\circ},20^{\circ})$     & 2.1867e+3 & 0.5742e+3 &  0.0605e+3 \\
   $(0^{\circ},40^{\circ})$     & 2.2902e+3 & 1.5089e+3 &  0.0436e+3 \\
   $(0^{\circ},60^{\circ})$     & 2.2261e+3 & 1.5380e+3 &  0.0462e+3 \\
   $(0^{\circ},80^{\circ})$     & 2.1598e+3 & 1.4089e+3 &  0.0513e+3 \\
  \hline
\end{tabular}
\end{table}

\section{Conclusions}\label{conc}

We have introduced a Bayesian joint separation and estimation
method for astrophysical images. The method is based on a Monte
Carlo technique and gives better reconstruction in the pixel
domain and frequency domain than two competitor methods. The
algorithm works quite well at high latitudes. If we approach the
galactic plane, the estimation results get worse. Especially at
the galactic plane, we have obtained the worst results, although
we have used a different initialization strategy.

Our new goal is the application of the proposed algorithm to
whole-sky maps. To avoid the difficulties inherent in this
problem, we plan to use the "nested numbering" structure provided
by the HEALPix \citep{Gorski05} package. In this format, we can
reach the indexes of the eight neighbors of each pixel on the
sphere. To calculate the pixel differences, we will implement a
gradient calculation method on the sphere by taking the
non-homogeneous spatial distances between the pixels on the sphere
into consideration.

\section*{Acknowledgments}

The authors would like to thank Anna Bonaldi,(INAF, Padova,
Italy), Bulent Sankur, (Bogazici University, Turkey) and Luigi
Bedini (ISTI, CNR, Italy) for valuable discussions. The simulated
source maps are taken from the Planck Sky Model, a set of maps and
tools for generating realistic Planck simulations made available
thanks to the efforts of the Planck Working Group 2 (WG2) team.
Some of the results in this paper have been derived using the
HEALPix \cite {Gorski05} package.

Koray Kayabol undertook this work with the support of the "ICTP
Programme for Training and Research in Italian Laboratories,
Trieste, Italy, through a specific operational agreement with
CNR-ISTI, Italy. Partial support has also been given by the
Italian Space Agency (ASI), under project COFIS (Cosmology and
Fundamental Physics). The project is partially supported by
CNR-CSIC bilateral project no: 2008IT0059.

\appendix

\section[]{Algorithm} \label{ALS}

One cycle of Adaptive Langevin Sampler for source separation. The
symbol $\longleftarrow$ denotes analytical update, the symbol
$\sim$ denotes update by random sampling.

%\begin{table}
%\caption{}
  %\centering
  %{{\parbox{10.0cm}{
  Find the initial mixing matrix with FDCCA \cite{Bedini07}.\\
  Find the initial source images using the LS solution.\\
  Initialize the parameters $\alpha_{l,d}^{0}$, $\beta_{l,d}^{0}$ and $\delta_{l,d}^{0}$\\
  for all source images, $l=1:L$
  \begin{quote}
    for all directions, $d=1:D$
    \begin{quote}
        $\langle\nu_{l,d}\rangle \longleftarrow
        \frac{N+\beta_{l,d}^{k}}{\beta_{l,d}^{k}} \left(1+\frac{\phi_{d}(\mathbf{s}_{l}^{k},\alpha_{l,d}^{k})}{\beta_{l,d}^{k}\delta_{l,d}^{k}}\right)^{-1}$\\
        $\alpha_{l,d} \longleftarrow
        \frac{\mathbf{s}_{l}^{T}\mathbf{G}_{d}^{T}\mathbf{s}_{l}}{\mathbf{s}_{l}^{T}\mathbf{G}_{d}^{T}\mathbf{G}_{d}\mathbf{s}_{l}}$\\
        $\delta_{l,d} \longleftarrow
            \langle\nu_{l,d}\rangle\frac{\phi_{d}(\mathbf{s}_{l},\alpha_{l,d})}{N}$\\
        $\beta_{l,d} \longleftarrow_{0}  [- \psi_{1}(\beta_{l,d}/2) + \log\beta_{l,d} + \langle\log\nu_{l,d}\rangle - \langle\nu_{l,d}\rangle +1 =
        0]$
    \end{quote}
    $\mathbf{w}_{l} \sim
        \mathcal{N}(\mathbf{w}_{l}|0,\mathbf{I})$\\
    $\overline{\mathcal{H}}(\mathbf{s}_{l}^{k}) \longleftarrow \mathrm{diag}
        \left\{\mathcal{H}(\mathbf{s}_{l})
        \right\}_{\mathbf{s}_{l} \longleftarrow
        \mathbf{s}_{l}^{k}}$\\
    $\mathbf{D} \longleftarrow 2 [\langle
    \overline{\mathcal{H}}(\mathbf{s}_{l}^{k})\rangle]^{-1}$\\
    $\mathbf{g}(\mathbf{s}_{1:L}^{k}) \longleftarrow [\nabla_{\mathbf{s}_{l}} E(\mathbf{s}_{l})]_{\mathbf{s}_{l}
        = \mathbf{s}_{l}^{k}}$\\
    produce $\mathbf{z} \longleftarrow \mathbf{s}_{l}^{k} - \frac{1}{2}\mathbf{D}\mathbf{g}(\mathbf{s}_{1:L}^{k})
    + \mathbf{D}^{\frac{1}{2}} \mathbf{w}_{l}$ from
    (\ref{lanvegin}).\\
    apply threshold to $\mathbf{z}$\\

    for all pixels, $n=1:N$
    \begin{quote}
        calculate $\varphi(z_{n},s_{l,n}^{k})$\\
    if $\varphi(z_{n},s_{l,n}^{k})\geq 1$ then $s_{l,n}^{k+1}=z_{n}$\\
          else produce $u\sim U(0,1)$.
    \begin{quote}
        if $u<\varphi(z_{n},s_{l,n}^{k})$ then $s_{l,i}^{k+1}=z_{n}$,\\
        else $s_{l,i}^{k+1}=s_{l,i}^{k}$
       \end{quote}
    \end{quote}
  \end{quote}
  for all elements of the mixing matrix, $(k,l)=(1,1):(K,L)$
  \begin{quote}
  $a_{k,l}  \longleftarrow  \frac{1}{\mathbf{s}_{l}^{T}\mathbf{H}_{k}^{T}\mathbf{H}_{k}\mathbf{s}_{l}} \mathbf{s}_{l}^{T}\mathbf{H}_{k}^{T}(\mathbf{y}_{k} - \mathbf{H}_{k} \sum_{i=1, i\neq
    l}^{L} a_{k,i}\mathbf{s}_{i}) u(a_{k,l})$
  \end{quote}
    %}}}
%\end{table}

\end{document}